\documentclass[reprint,
superscriptaddress,
longbibliography,
]{revtex4-2}

\usepackage{epsfig,amsmath,amssymb,bm,epsf,psfrag,bbm}
\usepackage{graphicx}
\usepackage{dcolumn}
\usepackage{bm}
\usepackage{amsbsy}
\usepackage{amsthm}
\usepackage{color, comment, verbatim}
\usepackage{ragged2e}
\usepackage{subfigure}
\usepackage{enumerate}
\usepackage{chngcntr}
\usepackage[T1]{fontenc}
\usepackage[english]{babel}
\usepackage[utf8]{inputenc}

\usepackage{dsfont}

\usepackage{tikz}
\usepackage[toc,page]{appendix}
\usepackage{titlesec}

\usepackage[colorlinks,breaklinks]{hyperref}
\usepackage{xcolor}

\makeatletter
\newcommand\org@hypertarget{}
\let\org@hypertarget\hypertarget
\renewcommand\hypertarget[2]{%
  \Hy@raisedlink{\org@hypertarget{#1}{}}#2%
  }
\makeatother

\hypersetup{
	bookmarksnumbered,
	pdfstartview={FitH},
	citecolor={darkgreen},
	linkcolor={darkred},
	urlcolor={darkblue},
	pdfpagemode={UseOutlines}}
\definecolor{darkgreen}{RGB}{50,190,50}
\definecolor{darkblue}{RGB}{0,0,190}
\definecolor{darkred}{RGB}{238,0,0}

\usepackage{booktabs}
\usepackage{tablefootnote}
\usepackage{threeparttable}

\newcommand{\ket}[1]{\ensuremath{\left|\right.\!{#1}\!\left.\right\rangle}}
\newcommand{\bra}[1]{\ensuremath{\left\langle\right.\!{#1}\!\left.\right|}}

\newcommand{\tr}{\textnormal{Tr}}

\begin{document}

\title{Classification of quantum states of light using random measurements through a multimode fiber}

\author{Saroch Leedumrongwatthanakun}
\email{saroch.l@psu.ac.th}
    \affiliation{Laboratoire Kastler Brossel, ENS-PSL Research University, CNRS, Sorbonne Universit\'e, Coll\`ege de France, 24 rue Lhomond, Paris 75005, France}
    \affiliation{Institute of Photonics and Quantum Sciences, Heriot-Watt University, Edinburgh, UK}
    \affiliation{Division of Physical Science, Faculty of Science, Prince of Songkla University, Hat Yai, Songkhla 90110, Thailand}
    
\author{Luca Innocenti}
    \affiliation{Universit\`a degli Studi di Palermo, Dipartimento di Fisica e Chimica – Emilio Segr\`e, via Archirafi 36, I-90123 Palermo, Italy}
    \affiliation{Centre for Quantum Materials and Technologies, School of Mathematics and Physics, Queen’s University Belfast, BT7 1NN Belfast, United Kingdom}
    
\author{Alessandro Ferraro}
    \affiliation{Centre for Quantum Materials and Technologies, School of Mathematics and Physics, Queen’s University Belfast, BT7 1NN Belfast, United Kingdom}
    \affiliation{Dipartimento di Fisica “Aldo Pontremoli,” Universit\`a degli Studi di Milano, I-20133 Milano, Italy}
    
\author{Mauro Paternostro}
    \affiliation{Universit\`a degli Studi di Palermo, Dipartimento di Fisica e Chimica – Emilio Segr\`e, via Archirafi 36, I-90123 Palermo, Italy}
    \affiliation{Centre for Quantum Materials and Technologies, School of Mathematics and Physics, Queen’s University Belfast, BT7 1NN Belfast, United Kingdom}
    
\author{Sylvain Gigan}
    \affiliation{Laboratoire Kastler Brossel, ENS-PSL Research University, CNRS, Sorbonne Universit\'e, Coll\`ege de France, 24 rue Lhomond, Paris 75005, France}

\begin{abstract}
Extracting meaningful information about unknown quantum states without performing a full tomography is an important task. Low-dimensional projections and random measurements can provide such insight but typically require careful crafting. In this paper, we present an optical scheme based on sending unknown input states through a multimode fiber and performing two-point intensity and coincidence measurements. A short multimode fiber implements effectively a random projection in the spatial domain, while a long-dispersive multimode fiber performs a spatial and spectral projection. We experimentally show that useful properties --- i.e., the purity, dimensionality, and degree of indistinguishability --- of various states of light including spectrally entangled biphoton states, can be obtained by measuring statistical properties of photocurrents and their correlation between two outputs over many realizations of unknown random projections. Moreover, we show that this information can then be used for state classification.
\end{abstract}

\maketitle

The evolution of a system of interest through an {uncharacterized quantum} channel transforms the state in an undesirable and seemingly detrimental way. Examples include unknown rotation of polarization through optical fibers, optical aberrations induced by atmospheric turbulence, and fluctuations of magnetic and electric fields in systems of trapped ions and cold atoms. Random matrix theory (RMT), which was originally developed to understand distributions of energy level spacings of heavy nuclei~\cite{Wigner1967}, can be used to successfully describe the statistical features arising from random evolutions. The breadth of its applications covers many areas of physics~\cite{Beenakker1997,Guhr1998,Collins2015}, all the way to information processing, where RMT has been insightfully applied to address compressed sensing~\cite{Candes2006}, random features of large-scale kernel machines~\cite{Rahimi2007}, and randomized algorithms for very large matrices~\cite{Mahoney2011}. In quantum physics,  randomized measurements -- a method of extracting information about the system of interest by performing measurements in random bases drawn from a certain ensemble -- have benefited from the use of RMT. Such an approach has been applied for the detection of entanglement in many-body systems without sharing reference frames between parties~\cite{Schlienz1995,Laskowski2012,Tran2015,Brydges2019,Ketterer2020,Knips2020momentrandom,Shuheng2022,Shuheng2022,Wyderka2022a}, the verification of quantum devices~\cite{Emerson2005,Elben2019,Eisert2020,Helsen2020}, and the simplification of state tomography~\cite{Gross2010b,Oren2017,Titchener2017,Wang2018,DeSantis2017,Banchi2018a}, among other tasks, thus consolidating its usefulness in estimating state properties with only a few copies~\cite{VanEnk2012,Dimic2018,Saggio2019,Huang2020}. This is not only remarkable but also a very valuable tool for the grounding of approaches to the characterization of quantum states and processes that do not rely on fully tomographic methods. The latter being typically very resource-expensive, set a {\it de facto} severe constraint to the scaling up of quantum technologies and their validation.

In optics, the propagation of light through scattering media or random interferometric processes is well described via RMT~\cite{Beenakker2009a,Rotter2017}, and results in complicated intensity patterns, known as speckles~\cite{Goodman2005}. They can be observed in interferograms of high-order intensity correlations, referred to as \textit{coincidence speckle}, and embody an interesting effect of optical coherence~\cite{Glauber1963,Saleh2000} that stems from the interplay of interference, indistinguishability, and correlations. Over the last two decades, substantial endeavours have been devoted to understanding the evolution of non-classical lights in complex scattering processes~\cite{Beenakker2009a,Ohad2022}. A wide range of topics has been theoretically investigated, for instance, the degradation of entanglement due to truncated and multimode detections~\cite{Aiello2004,VanVelsen2004,Puentes2007,Cande2014, Sorelli2020}, the transport of quantum noise~\cite{Patra2000b,Lodahl2005a,Lodahl2006,Lodahl2006a,Skipetrov2007}, and the dynamics of photon statistics in disordered or structured medium~\cite{Bromberg2009,Lahini2010,Carolan2014,EsatKondakci2016}. However, only a modest number of experiments have been carried out in this area~\cite{Lodahl2005,Puentes2007,Smolka2009,Peeters2010,VanExter2012,DiLorenzoPires2012}. An intriguing prediction is the presence of spatial intensity correlation averaged over many settings of random measurements~\cite{Lodahl2005,Smolka2009,Smolka2011,Ott2010}. The spatial intensity correlations in two-fold coincidence speckles result from both classical and quantum origins~\cite{Cande2013,Starshynov2016, Rigovacca2016} and depends upon both the scattering properties of a medium and the state of the incident light~\cite{Lodahl2005,Smolka2009,Ott2010,Cherroret2011,Starshynov2016,Walschaers2016,Li2019c,Phillips2019}.

The statistical distribution of two-fold coincidence speckles for a spatially maximally entangled biphoton state has hitherto been applied to predict purity and entanglement dimensionality~\cite{Beenakker2009}. An experimental demonstration based on the use of rotating diffusers as reprogrammable random transformations has been reported in Ref.~\cite{DiLorenzoPires2012}. The statistical moments of two-point spatial correlations have been recently extended to certify the degree of indistinguishability of many-particle quantum interference in the context of boson sampling problem~\cite{Walschaers2016,Walschaers2016a,Phillips2019}. This has been experimentally applied to classify the indistinguishability of three-photon interference on seven-mode integrated photonic waveguides~\cite{Giordani2018}. The statistical distribution of two-point spatial correlation~\cite{Phillips2019} has been used to validate the quantum interference of 50 indistinguishable single-mode squeezed states on a 100-mode phase-stabilized interferometer by ruling out the plausible hypotheses of outcome stemming from thermal states and distinguishable photons~\cite{Qin2020}.

In this work, we propose a simple optical implementation of a reconfigurable random transformation obtained by randomly sending a state of light through a multimode fiber using a spatial light modulator.
By changing the length of the fiber to increase dispersions, this implementation allows exploiting, besides the polarization and spatial degrees of freedom, also the temporal modes of light. We employ the proposed apparatus to demonstrate the use of random photocurrent measurements, second-order intensity correlations (two-fold coincidences), and normalized second-order intensity correlation between two truncated output modes, to probe the properties of input states for the purposes of state classification. Various paradigmatic input states, such as one- and two-photon states, N00N states, and spectrally entangled biphoton states are used in our experimental demonstration.

We observe the flattening of the statistical distribution of normalized second-order intensity correlation due to the presence of two-photon interference, which indicates a good measure of the degree of indistinguishability as compared to that computing from the unnormalized two-fold coincidences widely used in previous seminal demonstrations. The study of statistics of normalized and unnormalized second-order intensity correlation is also extended to the presence of spectral-temporal entanglement and dispersions. 

We demonstrate that the use of random measurements on the three types of detected signals reveals useful distinct statistical signatures. By jointly analyzing statistical moments of these outcomes, we resolve the properties of the input states, including the purity, dimensionality, and indistinguishability, which are then used to classify the states without the need to perform full state tomography. The simplicity of the setting put forward in this implementation as well as our study on the accuracy in characterizing the state properties shows the potential of the proposed experiment to embody a valuable tool in the quest for a resource-effective classification of quantum states of light.

\begin{figure*}[htp]
	\centering
	\includegraphics[width=2\columnwidth]{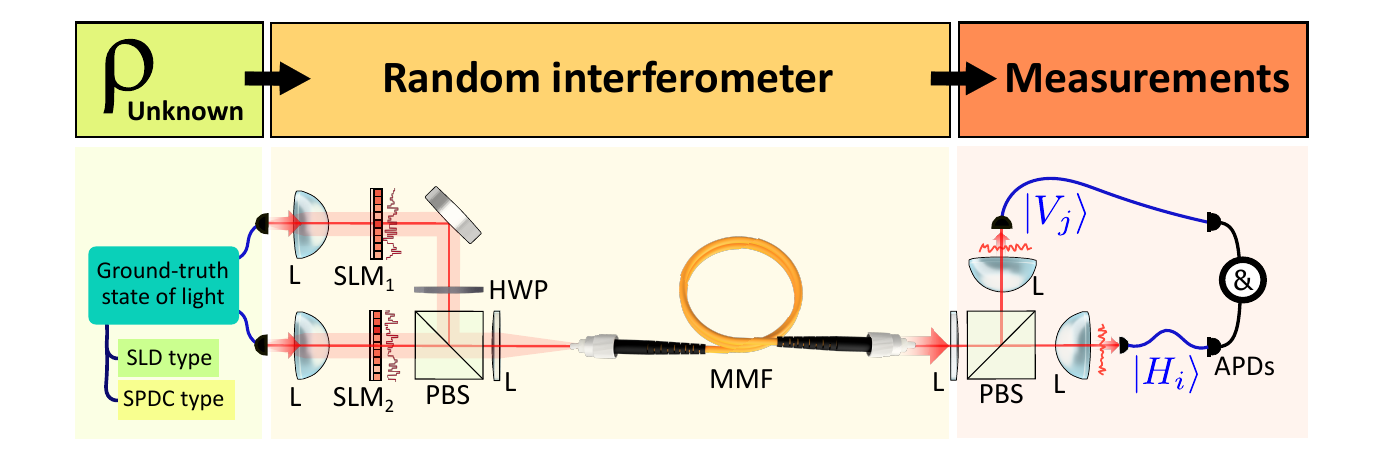}
	\caption{\textbf{Concept and experimental scheme}: An unknown state of light $\rho_{\text{unknown}}$ is evolved through a random interferometer and mapped onto a high-dimensional output that is subjected to a measurement step. The statistical features of the outcomes are used to infer properties of the state. In the experiment, ground-truth states of light, as listed in~Sec.\ref{sec:Ground-truth states}, have been generated through Spontaneous Parametric Down-Conversion (SPDC) and Superluminescent Diode (SLD). Such states are randomly launched into a multimode fiber (MMF) using the spatial light modulators $\text{SLM}_1$ and $\text{SLM}_2$ that are placed on the Fourier plane of two input orthogonal polarization channels of the MMF. At each setting of random measurement, randomly generated holograms are displayed on the SLMs, and a state is controlled and evolves through the MMF. We probed the output states on randomly selected two-mode subspaces assigned at two different diffraction-limited spots on the near-field plane of the MMF and associated with orthogonal polarizations, labeled $\ket{H_i}$ and $\ket{V_j}$. Photocurrents ($I$), two-fold coincidence counts ($C$), and normalized second-order intensity correlation ($g^{(2)}$) are measured by avalanche photodiodes (APDs) and a coincidence electronic circuit with the coincidence window of 2.5 ns. (L: lenses, HWP: half-wave plate, PBS: polarizing beamsplitter).}
	\label{fig1}
\end{figure*}

\section{Implementation of random measurement}
\label{sec:Implementation of random measurement}
We implement random transformations by evolving input states through a spatial light modulator (SLM) and a graded-index multimode fiber (MMF), acting jointly as a random interferometer~(Fig.~\ref{fig1}).
Many random transformations can be realized by displaying different phase patterns on the SLM, which controls light coupling through the MMF propagating onto the large output space.
We here consider the scenario where an input state evolves through the random interferometer, and therefore the dimension of the input state is smaller than the total number of propagating modes of the MMF.
Moreover, we study the case where the random measurement is performed only on two fixed output modes, thus establishing our approach as being of a constrained-resource nature.
Since the random interferometer is sufficiently large, unitary, and random, the random transformation on two truncated output spaces can be considered to be efficiently drawn from an independent and identically distributed complex Gaussian random matrix~\cite{Zyczkowski2000}.
This means that correlations and unitary constraint presented in a sub-part transmission matrix of the MMF linking to two truncated outputs is negligible, provided that the number of truncated modes $(p=2)$ satisfies $p\leq m^{1/6}$~\cite{Aaronson2011a} with $m\sim400$ the {number of modes propagating across the optical fiber}.
The statistical property of the random transformation onto two truncated outputs is provided in the supplementary information accompanying this manuscript~(\ref{SI:Statistical analysis of random measurements}). 

For a given measurement setting, photocurrents at two outputs $I_1$ and $I_2$, and two-fold coincidence counts $C$ are measured by threshold non-resolved photon-number detectors within a coincidence window $\tau_{C}$. The normalized second-order intensity correlation $g^{(2)}$ is determined as $g^{(2)}=C/R$, where we have introduced the accidental coincidences $R=\tau_{C} I_1 I_2$. The details of the experimental setting are provided in~\ref{SI:Experimental parameters and status of ground-truth light sources}.

\section{Ground-truth states} 
\label{sec:Ground-truth states}

We test the ability of our apparatus to perform state classification with a set of states spanning one and two spatial modes. 
The first state we consider is a mixture of two spatial modes generated by the amplified spontaneous emission of a superluminescent diode (SLD) passing through a $810\pm1$ nm bandpass filter. Such incoherent radiation is used as a benchmark classical state to validate our approach. The remaining ground-truth states are then prepared from type-II spontaneous parametric down-conversion (SPDC) by pumping a periodically poled potassium titanyl phosphate crystal (ppKTP) with a 405-nm continuous-wave laser in the single spatial mode generation. The correspondingly generated spatially separable biphoton states are entangled in frequency (cf.~\ref{SI:SPDC_simulation}). The degree of indistinguishability of the ground-truth biphoton states is controlled by adjusting the temporal delay between signal and idler photons.
A two-photon Fock state and an $N=2$ N00N state~\cite{Lee2002} $(\ket{20}+\ket{02})/\sqrt2$ are generated using Hong–Ou–Mandel (HOM) interference~\cite{Hong1987}.
The HOM visibilities of the sources before and after the experiment are provided in~\ref{SI:Experimental parameters and status of ground-truth light sources}. Finally, a single-photon Fock state is generated through a standard heralded approach.

By varying the lengths of the MMF, we can spatially control the temporal mixing of the interferometer, therefore being able to produce both narrowband and broadband states.
For the bandwidth of our sources, a length of 55 cm and 25 m correspond to narrowband (non-dispersive) and broadband (dispersive) regimes, respectively. In the narrowband regime, all states were tested.
In the broadband regime, where random different spatial modes allow to be mapped to the temporal domain, only SLD source, indistinguishable, and distinguishable biphoton states, were studied.
Our experimental characterization has also shown that the HOM visibility of the sources is preserved after propagating through the 55-cm long MMF, and persists when using the 25-m long fiber.

\section{Results and discussions}
\label{Results and discussions}
 Statistical moments and distributions of outcomes from random measurements can be used to infer various properties of the unknown states. Building on the theoretical framework presented in~\ref{SI:Theoretical frameworks}, here we demonstrate the use of statistical markers to achieve quantum state classification.
 In particular, in Sec.~\ref{sec:estimation of state properties by random measurements}, we demonstrate recovery of the number of occupied modes, purity, and entanglement dimensionality for input maximally entangled two-photon states. We then report the effect of quantum interference on the normalized second-order correlation function and propose a figure of merit for measuring the degree of indistinguishability based on its first two statistical moments (see also Sec.~\ref{sec:On the effect of quantum interference and indistinguishability}). The figure of merit is extended to the presence of dispersion and frequency entanglement as discussed in Sec.~\ref{sec:On the frequency-entangled biphoton states}. Finally, an example of state classification based on such properties is given in Sec.~\ref{sec:State classification}.

\begin{figure*}[htp]
	\centering
	\includegraphics[width=\linewidth]{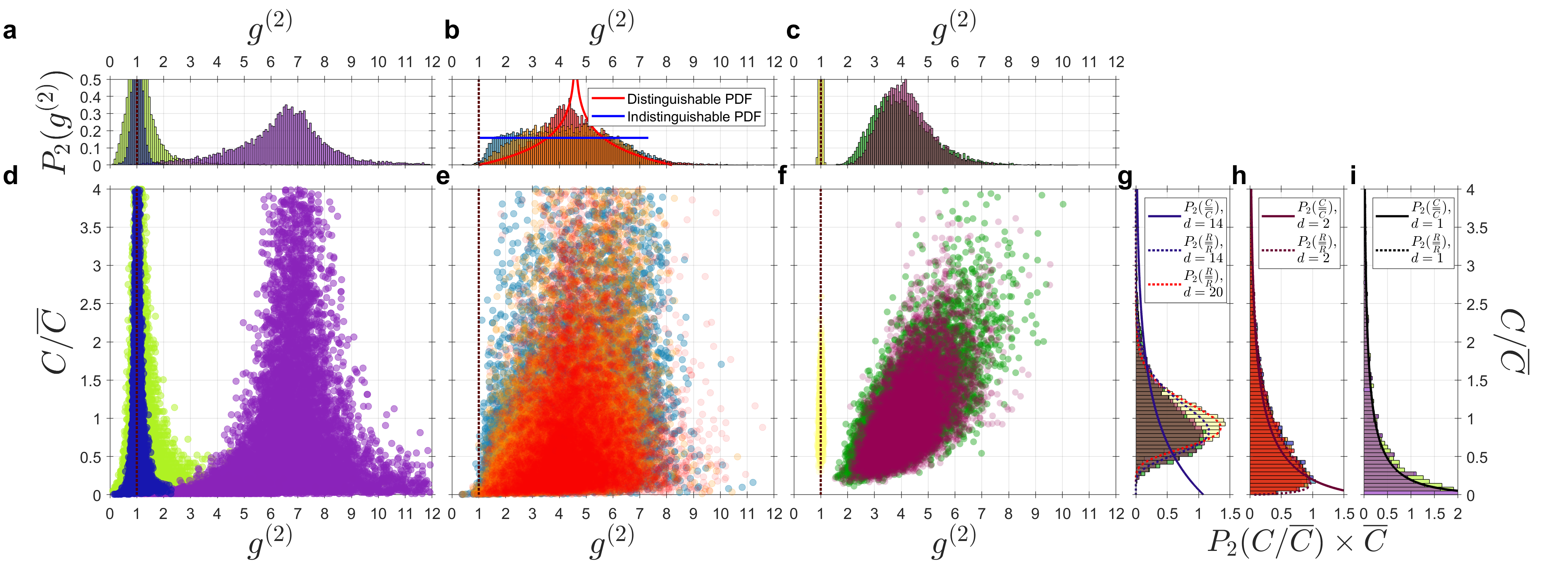}
	\caption{\textbf{Statistics of two-fold coincidences}: \textbf{(a-c)} Statistical distributions of experimental normalized second-order correlation $P_2(g^{(2)})$ for \textbf{(a)} heralded single-photon state (light green), incoherent source (dark blue) and the two-photon Fock state $\ket{2}$ (magenta) in the 55-cm long MMF. The first two states have the means at the accidental coincidence (the red dashed line, $\overline{g^{(2)}}=1$), while the two-photon Fock state shows $\overline{g^{(2)}}>1$. \textbf{(b)} $P_2(g^{(2)})$ for the group of two-mode states in a 55-cm long MMF: Indistinguishable biphoton state (blue), N=2 N00N state (orange), and distinguishable biphoton state (red). Their distributions are different owing to the presence of quantum interference. The histogram of indistinguishable biphotons shows the flat distribution as predicted by the probability density functions (PDF) $P_2(g^{(2)})$, represented by the blue solid line. The red curve represents the prediction for distinguishable biphotons, Eq.~\ref{eq:PDFg2Dismain}. \textbf{(c)} $P_2(g^{(2)})$ for the group of states evolving through the 25-m long MMF: Incoherent source (light yellow), indistinguishable (green) and distinguishable (light magenta) biphoton states. The incoherent state has the means at $\overline{g^{(2)}}=1$, while for the biphoton states ($g^{(2)}>1$) the width of the indistinguishable case is broader than that of the distinguishable case. \textbf{(d-f)} Correlation of two-fold coincidences $C/\overline{C}$ and normalized second-order correlation $g^{(2)}$ for the ground-truth states as previously labelled. Each circle on the scatter plots displays each outcome of random measurements. \textbf{(g-i)} Statistical distributions of two-fold coincidences $P_2(C/\overline{C})$ for the corresponding ground-truth states. The solid curves indicate the PDF of two-fold coincidences $P_2(C/\overline{C})$ for pure $d$-dimensional spatially maximally entangled biphoton states, Eq.~\ref{eq:PDF_Cmain}. The dashed curves represent the PDF of accidental coincidences $P_2(R/\overline{R})$, Eq.~\ref{eq:PDF_ACCmain}. \textbf{(g)} $P_2(C/\overline{C})$ for the group of states propagating through 25-m MMF: Indistinguishable and distinguishable biphoton states and incoherent state. The first two biphoton states show the same distribution which cannot be described by the PDF of expected accidental coincidences $P_2(R/\overline{R})$ for $d\approx14$ predicted from the visibility of intensity. This is in contrast to the distribution of the incoherent source that is classically well predicted from the visibility of intensity with $d=20$. \textbf{(h)} $P_2(C/\overline{C})$ for the group of two-mode states propagating through the 55-cm long MMF: Indistinguishable biphoton state, N=2 N00N state, distinguishable biphoton state, and incoherent source, presents no statistical difference between their distributions. \textbf{(i)} $P_2(C/\overline{C})$ for the group of single-mode states: single-photon and two-photon Fock states show the same distribution of $C/\overline{C}$ with $d=1$.}
	\label{fig:Mainresults}
\end{figure*} 

\subsection{Estimation of state properties by random measurements}
\label{sec:estimation of state properties by random measurements}

\noindent
\textit{Number of occupied modes ---}
The first useful property of the states that can be used for the classification is the number of occupied modes $(d)$, which can be predicted by using the statistical properties of the measured photocurrent $(I)$, typically via the calculation of the intensity visibility, $\mathcal{V}_{I}:={\operatorname{Var}\left(I\right)}/{\overline{I}^{2}}=1/d$. In passing, we mention that $\sqrt{\mathcal{V}_{I}}$ is referred to as \textit{speckle contrast} in literature~\cite{Goodman1976,Goodman2005}. The probability density function (PDF) of intensity $P_1(I/\overline{I})$ and the estimated values of $d$ well describe all types of ground-truth states, as reported in Fig.~\ref{fig:PDF_I1}. It is for instance well known that a single coherent mode will result in a contrast of 1, while the incoherent addition of $d$ incoherent modes results in a contrast of $1/\sqrt{d}$~\cite{Goodman1976,Goodman2005}. As this is only the first order intensity correlation, it cannot be used to distinguish a classical state from a non-classical one; it provides only information about the number of modes that the state occupies. In general, additional information on a state can be extracted by measuring outcomes in high-order intensity correlation functions~\cite{Glauber1963a}. In the case of random measurements, inferring the information of a state from the statistical properties of outcomes from unknown measurements is possible~\cite{Beenakker2009} and the generalization of the theory is still under development~\cite{Brunner2021}.

\noindent
\textit{Purity and entanglement dimensionality ---}
By incorporating measurements of the second-order intensity correlation function, one can obtain the purity $\mathcal{P}$ of the density matrix $\rho$ of arbitrary monochromatic biphoton states from the first two statistical moments of intensity $I$ and two-fold coincidence $C$~\cite{Beenakker2009,VanEnk2012} as $\mathcal{P}=\tr \rho^2 =\mathcal{V}_{C}-2 \mathcal{V}_{I}$, where $\mathcal{V}_{C}$ is the visibility of two-fold coincidence $\mathcal{V}_{C}:={\operatorname{Var}(C)}/{\overline{C}^{2}}$. In the case of the pure monochromatic maximally entangled biphoton state, the corresponding entanglement dimensionality ($D=d/2$) can then be estimated by $\mathcal{V}^{\text{pure}}_{C}=1+1/D$, and the PDF of two-fold coincidence $P_2(C/\overline{C})$ has the analytic form of the K-distribution, which reads
\begin{equation}
	\label{eq:PDF_Cmain}
	P_{2}\left(\frac{C}{\overline{C}}\right)=\frac{2d}{\Gamma(d)}\left(d \frac{C}{\overline{C}}\right)^{\frac{d-1}{2}} \mathcal{K}_{d-1}\left(2 \sqrt{d \frac{C}{\overline{C}}}\right),
\end{equation}
where $\Gamma$ is the gamma function and $\mathcal{K}_{d-1}(x)$ is a modified Bessel function of the second kind. 

In the experiment, we investigated the latter cases of $d=2$ monochromatic biphoton states. The results of the measured visibilities and purity are reported in Table~\ref{table:PandDtableApp} and the distributions are shown in Fig.~\ref{fig:Mainresults}h for indistinguishable biphoton state, N=2 N00N state, and distinguishable biphoton state, which present very similar feature.
Besides, they almost overlap with the distribution of the incoherent source that originates from accidental coincidence $P_2(R/\overline{R})$, which is
\begin{equation}
	\label{eq:PDF_ACCmain}
	P_{2}(\frac{R}{\overline{R}})=\frac{2}{\Gamma(d)^2}d^{2d} \left(\frac{R}{\overline{R}}\right)^{d-1} \mathcal{K}_{0}\left(2d \sqrt{\frac{R}{\overline{R}}}\right).
\end{equation}
This thereby results in the difficulty in the classification task on these states using solely the statistics of $C$, albeit the ideal simulation results indicate the different distribution of the three cases at $C$ close to 0 (as presented in Fig.~\ref{figPDFC_sim}a). The experimental distribution of the indistinguishable biphoton state exhibits less probability of detecting coincidence counts close to zero than that predicted by the K-distribution. We note that this effect was also present in the experimental data for high-dimensional entangled biphoton states~\cite{DiLorenzoPires2012}. We attribute this effect to the contribution of noise sources in each setting of random measurement, including dark counts, accidental coincidences, and finite exposure time which usually causes a broadening of the distribution. As further discussed in ~\ref{sec:Estimation of the purity}, the contribution of noise sources might be interpreted as a classical mixture of many pure biphoton states. In addition, the long tail of K-distribution also results in an unreliable estimation of $\mathcal{V}_{C}$. Both give rise to the underestimation of the measured purity (cf.~Table~\ref{table:PandDtableApp}). The use of statistical properties of the coincidences $C$ thus has a practical limitation in the estimation of purity at low $d$ using only the first two statistical moments.

\subsection{On the effect of quantum interference and indistinguishability}
\label{sec:On the effect of quantum interference and indistinguishability}
Imperfect indistinguishability between photons affects quantum interference. One may think that the quantum interference might have signatures in the statistical distribution of two-fold coincidences $P_2(C/\overline{C})$. Unfortunately, as reported above (Fig.~\ref{fig:Mainresults}h), this phenomenon is hardly observed in the experiment as compared to the ideal simulation (cf. ~Fig.~\ref{figPDFC_sim}a). Here, we demonstrate nevertheless that the effect of indistinguishability can be clearly unveiled from the statistical distribution of the normalized second-order correlation function $g^{(2)}$. As presented in Fig.~\ref{fig:Mainresults}b, in the cases of indistinguishable photons and an N=2 N00N state, we observed a broader and flatter statistical distribution compared to the one associated with distinguishable photons. The key reason is that the normalized second-order correlation function $g^{(2)}$ filters out the effect of the fluctuation of intensity speckles that are present in the two-fold coincidence speckles. The contribution of two-photon Hong-Ou-Mandel interference due to random projections, i.e., random varieties of HOM dips and peaks, therefore increases the variance of the distributions. The features are also evident in the simulated results (cf.~Sec.~\ref{SI:Theoretical frameworks}) supporting our observation where we found that the distribution is uniform for the indistinguishable case as compared to the symmetric negative-log distribution $P(g^{(2)}_\text{Dis.}/\overline{g^{(2)}})$ for distinguishable case,
\begin{equation}
    \label{eq:PDFg2Dismain}
	P\left(\frac{g^{(2)}_\text{Dis.}}{\overline{g^{(2)}}}\right)= -\frac{\log{|g^{(2)}_\text{Dis.}/\overline{g^{(2)}}-1|}}{\pi\overline{g^{(2)}}}.
\end{equation}

Moreover, the variance of the normalized second-order correlation for the N=2 N00N state is lower than that for the indistinguishable biphoton state. This arises from the fact that the N00N state experiences dephasing effects since the state is very sensitive to phase fluctuation at the inputs of the random interferometer that modulates faster than the exposure time in each random measurement. Consequently, $g^{(2)}$ detects only the root mean square response of the fast sinusoidal oscillation arising from interference originating from the use of a  N00N-state. We thus observed a $\sqrt{2}$ reduction of the variance for N00N states as compared to the case of the indistinguishable biphotons. The result is observed directly on the visibility of the normalized second-order correlation $\mathcal{V}_{g^{(2)}}:=\operatorname{Var}\left(g^{(2)}\right)/\overline{g^{(2)}}^{2}$ where the ratio of the measured visibility between the indistinguishable biphoton state and the N00N state is $\mathcal{V}^{\text{Indis.}}_{g^{(2)}}/\mathcal{V}^{\text{N00N}}_{g^{(2)}}=1.4\pm0.2\approx \sqrt{2}$. Therefore, the results demonstrated the effectiveness of the statistical properties of the normalized second-order correlation function in distinguishing the three types of biphoton states addressed herein.

\subsection{On the effect of dispersion on spectrally entangled biphoton states}
\label{sec:On the frequency-entangled biphoton states}
The next question of interest is to study the statistical outcomes of unknown states propagating through the 25-m long dispersive fiber. We launched various states of light, in particular spectrally entangled biphoton states, through different spatial modes of the MMF. The modal dispersion of the MMF causes an input finite pulse to temporally spread~\cite{Shen2005,Carpenter2014b}. In the spectral domain, this is equivalent to a narrow spectral response of the optical system, characterized by the spectral correlation bandwidth $\Delta\lambda_m$, in which a light source with bandwidth $\Delta\lambda_s$ will generate an incoherent sum of speckles over the number of temporal/spectral modes $N_{\lambda} = \Delta\lambda_s/\Delta\lambda_m$~\cite{Andreoli2015,Mounaix2016b}. The decrease in the outcome visibility of intensity speckles at the output of the long fiber can thus be used to estimate the number of occupied temporal/spectral modes. For spectrally entangled biphoton states, we measured the spectral incoherent modes of $\sim7$ which corresponds to the total number of occupied modes of $d\approx14$ given the two input modes of orthogonal polarizations (cf.~Table~\ref{table:PandDtableApp}).

In this situation, the spectral-temporal correlation of the SPDC source comes into play~\cite{Cherroret2011,Cande2013}. Two-photon quantum interference of spectrally entangled states is known to result in two types of dispersion cancellations: non-local and local~\cite{ Steinberg1992, Steinberg1992a, Franson1992, Franson2009}. At a given setting of a random interferometer, photons from biphoton pairs propagate and disperse differently through the fiber with a probability of being detected at two outputs. In the first case of non-local dispersion cancellation~\cite{Franson1992}, biphoton pairs propagate to separate paths and end up detected on two separate outputs, this results in the cancellation of the dispersion of one path to another path that has the opposite sign of dispersion, therefore maintaining their two-fold coincidences without the HOM interference. Whereas in the case of local dispersion cancellation~\cite{Steinberg1992}, the HOM interference can occur, biphoton pairs that experience different dispersions have a chance to interfere and exchange their wavefunctions (spatial-mode mixing, i.e., a possibility that signal and idler photons have a possibility to be detected on both outputs). Two types of paths contribute to such interference: one where the two paths of a biphoton pair are directly transmitted to the detectors, and one where they exchange paths. Only relative differences in dispersion between the paths can contribute to the distinguishability. Yet, the relative differences are partially compensated owing to frequency anticorrelation that erases which-path information~\cite{Steinberg1992a,Franson2009}, resulting in the robustness of quantum interference. As a result, we observed (Fig.~\ref{fig:Mainresults}c) that the statistical distribution of $g^{(2)}$ in the case of the indistinguishable biphoton state is broadened, compared to the case of the distinguishable biphoton state. The statistical distribution of $g^{(2)}$ is thus also useful for probing quantum interference in the presence of dispersion and spectral-temporal entanglement. It is noteworthy that the statistical signature of frequency entanglement was theoretically proposed but in terms of the means of coincidence rate, $\overline{C}/\overline{R}$~\cite{Cherroret2011,Cande2013}, in which our experiment cannot provide the conclusive result because of the requirement of the subtle control that is sensitive to the brightness of the light sources. In terms of the distributions, the theoretical prediction is unknown, to our knowledge.

Lastly, we compared the measurements of the normalized second-order correlation function $g^{(2)}/\overline{g^{(2)}}$ with the analogous two-fold coincidences $C/\overline{C}$ as depicted in Fig.~\ref{fig:Mainresults}(d-f). We observed that $g^{(2)}$ is clearly correlated to the two-fold coincidences $C$ in the case of the spectrally entangled states (Fig.~\ref{fig:Mainresults}f), whereas this is not the case for the other states addressed here (Figs.~\ref{fig:Mainresults}d and~\ref{fig:Mainresults}e). The positive correlation between the two indicators hence strengthens our claim in the previous section on the effect of the fluctuation in intensity speckles on the two-fold coincidence speckles since here the dispersion causes $P_1(I/\overline{I})$ to be a narrow distribution, i.e., low speckle visibility (cf.~\ref{fig:PDF_I1}c), while the dispersion cancellations can maintain the quantum interferences, hence, resulting in a high speckle visibility in the normalized second-order correlation function $\mathcal{V}_{g^{(2)}}$.

As for the aim for classification, the distributions of $C/\overline{C}$ in Fig.~\ref{fig:Mainresults}(g) are observed to be identical for different degrees of indistinguishability and also have a feature similar to the distribution of accidental coincidences at a lower $d$. Consequently, the state classification with information solely estimated from the statistics of two-fold coincidences $C$, which is commonly used in literature, is not sufficient, and taking into consideration the information from the $g^{(2)}$ distribution is beneficial both for the non-dispersive and dispersive cases.

\subsection{State classification}
\label{sec:State classification}
\begin{figure}[htp]
	\centering
	\includegraphics[width=\linewidth]{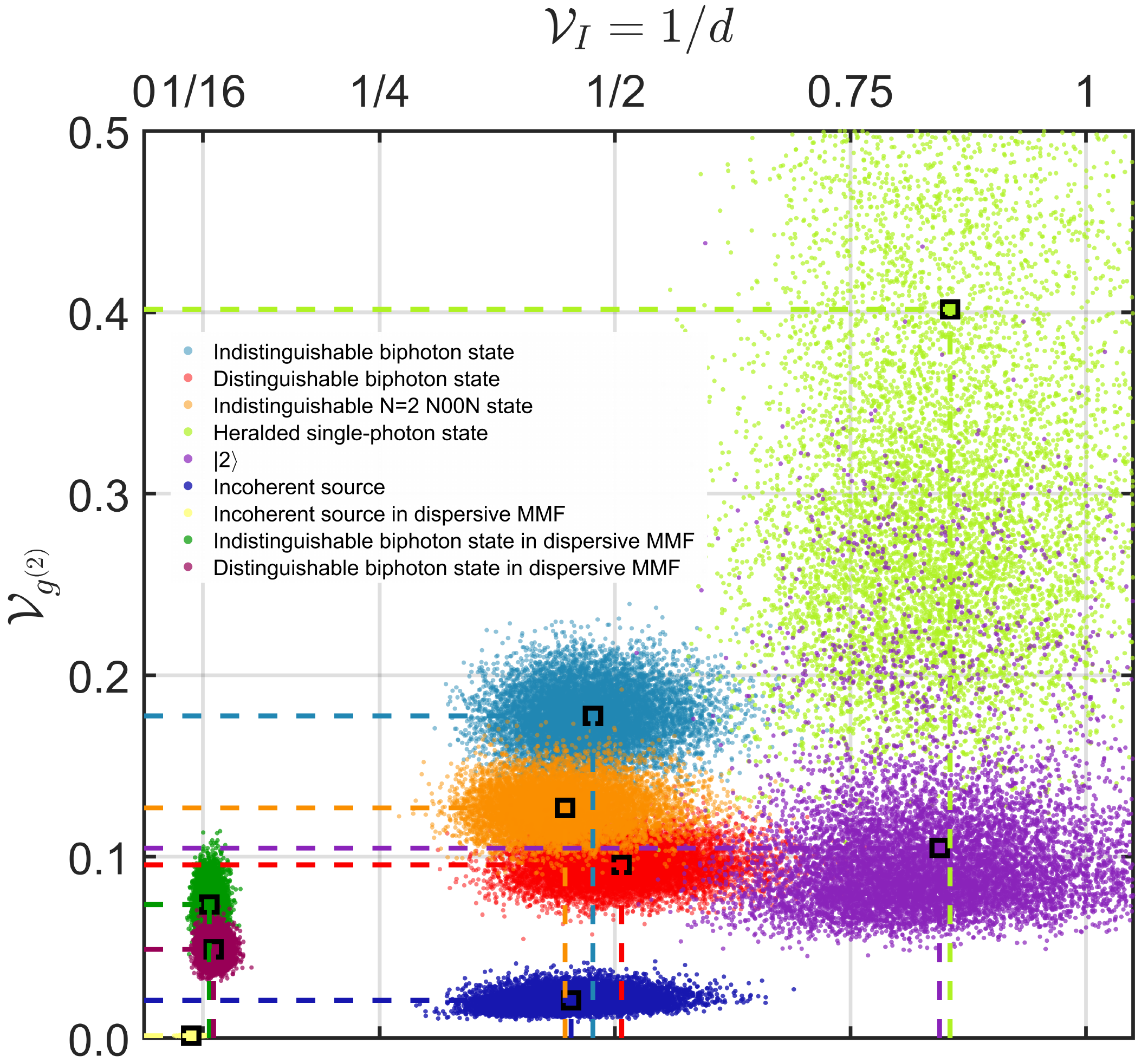}
    \caption{
    {\bf Classification of input states from intensity and second-order correlations}
    Visibility $\mathcal V_I$ and second-order correlations $\mathcal V_{g^{(2)}}$ for the different input states.
    The clustering patterns clearly show that the different classes of input states can be separated using only these two statistical features. In particular, it is possible to discriminate both the distinguishability and the number of occupied modes of the input states.
    }
	\label{fig:StateClassification}
\end{figure}

In this section, we demonstrate the possibility of classifying states based on their statistical properties.
More specifically, we are able to retrieve the number $d$ of occupied modes, as well as the distinguishability of the input photons, through information gathered from the visibility $\mathcal V_I$ and normalized second-order correlation $\mathcal V_{g^{(2)}}$ 
In Fig.~\ref{fig:StateClassification} we show how different classes of input states can be distinguished through such figures of merit. We present the visibilities estimated from $200$ random measurements performed on the same experimental input state.
Clearly, $\mathcal V_I$ and $\mathcal V_{g^{(2)}}$ can be used to independently probe the number of occupied modes and the distinguishability of the input states, respectively. The technique can also discriminate distinguishable from indistinguishable spectrally-entangled biphoton states. The overestimation of the number of occupied modes for the single-photon and two-photon Fock states is due to the presence of dark counts, an imperfection that can be corrected (cf.~Table~\ref{table:PandDtableApp}). The spreading of $\mathcal{V}_{g^{(2)}}$ values for the single-photon state results instead from accidental coincidences. Better classification performances could be achieved by incorporating the constraint $\overline{g^{(2)}}=1$, which would help discriminate single- and two-photon Fock states.

\section{Conclusions and outlook}
\label{Conclusion and outlook}
We have experimentally demonstrated the effectiveness of a state classifier based on the combined use of an SLM and a multimode fiber.
The scheme performs state classification using the statistical properties from an ensemble of random measurements. In particular, intensity and second-order intensity correlations are used to retrieve information about the number of occupied modes, purity, and indistinguishability of the input states, without the need to perform full state tomography.
The benefits and limitations of using the statistical properties of two-fold coincidences $C$ and normalized second-order correlation $g^{(2)}$ are investigated using different ground-truth states, including spectrally entangled biphoton states in the dispersive fiber.
The investigation of different classes of input states, having different statistical properties, is an interesting venue for future research.
In addition to the statistical properties reported here, the optical apparatus also offers the capability of implementing a reprogrammable linear optical circuit~\cite{Leedumrongwatthanakun2020} and can be seamlessly incorporated with an array of coincidence detectors, hence enabling the implementation of quantum multi-outcome measurements.~\cite{Bruschini2019,Goel2022,Makowski2023}. This allows the feasibility of performing optimal measurements, thus enabling a direct estimation of the properties of the state, e.g., purity, dimensionality, and type of entanglement. The possibility of implementing state tomography in high dimensions will be the subject of future investigations.

\section*{Acknowledgements}
 The work is supported by European Research Council (ERC) (724473). S.G. is a member of the Institut Universitaire de France (IUF). L.I. acknowledges support from MUR and AWS under project PON Ricerca e Innovazione 2014- 2020, “calcolo quantistico in dispositivi quantistici rumorosi nel regime di scala intermedia” (NISQ - Noisy, Intermediate- Scale Quantum). M.P. is supported by the European Commission through the Horizon Europe EIC Pathfinder project QuCoM (Grant Agreement No.\,101046973), the Leverhulme Trust through the Research Project Grant `Ultracold quantum thermo-machine' (UltraQuTe, grant number RGP-2018-266), MSCA co-funding of regional, national and international programmes (grant number 754507), the Royal Society Wolfson Fellowship (RSWF/R3/183013), the UK EPSRC (grant EP/T028424/1), the Department for the Economy Northern Ireland under the US-Ireland R\&D Partnership Programme. 


\newpage
\bibliographystyle{apsrev4-1fixed_with_article_titles_full_names}
\bibliography{Reference}

\hspace{5cm}
\newpage

\newpage
\onecolumngrid
\newpage
\begin{center}
\textbf{\large Supplementary information for: \\Classification of quantum states of light using random measurements through multimode fiber}\\[1cm]
\end{center}

\twocolumngrid
\renewcommand\thesection{SI.\arabic{section}}
\setcounter{section}{0}
\setcounter{equation}{0}
\setcounter{figure}{0}
\setcounter{table}{0}
\setcounter{page}{1}
\renewcommand{\theequation}{S\arabic{equation}}
\renewcommand{\thefigure}{S\arabic{figure}}
\renewcommand{\thetable}{S\arabic{table}}

\section{Theoretical frameworks and simulation}
\label{SI:Theoretical frameworks}

\subsection{Statistical distribution of intensity speckles}
\label{SI1a}
Classical speckles can be used to reveal characteristics of light forming it. Statistical properties of intensity speckles provide information about the first order degree of coherence. In the case of incoherent uniform mixing of coherent states, the probability density function (PDF) of intensity speckle is
\begin{equation}
	\label{eq:PDF_I}
	P_1\left(\frac{I}{\overline{I}}\right)=\frac{d^{d}}{\Gamma(d)} \left(\frac{I}{\overline{I}}\right)^{d-1} e^{-d {I}/{\overline{I}}},
\end{equation}
where $d$ represents a number of occupied mode, $\Gamma$ is the gamma function, and $\overline{I}$ is an average intensity. This PDF is known as a gamma density function of order $d$~\cite{Goodman2005}.

\subsection{Statistical property of two-photon speckles}
\label{SI1b}
The speckle pattern of two-fold coincidences generated from a biphoton state is known as two-photon speckle. Its statistical properties were analyzed in Ref.~\cite{Beenakker2009}. For a pure spatially maximally entangled biphoton state of dimension $d$, the PDF of the intensity speckle $P_1 (I)$ has the same form as in the case of a classical full-mixture of coherent states (Eq.\ref{eq:PDF_I}). And, the PDF of two-photon speckle $P_2 (C)$ reads,
\begin{equation}
	\label{eq:PDF_C}
	P_{2}\left(\frac{C}{\overline{C}}\right)=\frac{2d}{\Gamma(d)}\left(d \frac{C}{\overline{C}}\right)^{\frac{d-1}{2}} \mathcal{K}_{d-1}\left(2 \sqrt{d \frac{C}{\overline{C}}}\right),
\end{equation}
where $\mathcal{K}_{d-1}(x)$ is a modified Bessel function of the second kind. This distribution is known as K-distribution~\cite{Goodman2005} and has a classical analogy with an intensity speckle. It appears, for example, in the study of rogue waves~\cite{Arecchi2011} and shows a non-exponential long decay at low $d$~\cite{Abraham2010}.

For a pure monochromatic biphoton state, the PDF of two-photon speckle is fixed via the PDF of intensity speckle owing to the one-to-one mapping between $P_1 (I)$ and $P_2 (C)$~\cite{Beenakker2009}. $P_2 (C)$ does not provide additional information about a given pure biphoton state~\cite{Beenakker2009}. Nevertheless, the PDF of two-photon speckles can be used to distinguish a pure biphoton state from other states. For example, in the case of a single-photon state, no genuine two-fold quantum coincidences exist. Furthermore, in the case of a mixed biphoton state, no such one-to-one mapping between $P_1 (I)$ and $P_2 (C)$ exists~\cite{Beenakker2009}. The purity $\mathcal{P}$ of a biphoton density matrix $\rho$ can be directly obtained from the first two statistical moments of intensity speckle and two-fold coincidence speckle~\cite{Beenakker2009,VanEnk2012} as follows:
\begin{equation}
	\label{eq:Purity2009}
	\mathcal{P}:=\tr \rho^2 =\mathcal{V}_{C}-2 \mathcal{V}_{I},
\end{equation}
where $\mathcal{V}_{I}$ is the visibility of intensity speckles and $\mathcal{V}_{C}$ is the visibility of two-fold coincidence speckles defined as
\begin{equation}
	\label{eq:V_C}
	\mathcal{V}_{C}:=\operatorname{Var}(C)/\overline{C}^{2}.
\end{equation}

The entanglement dimensionality ($D=d/2$) of a pure spatially maximally entangled biphoton state is obtained from both visibilities,
\begin{subequations}
	\begin{align}
		&\mathcal{V}^{\mathrm{pure}}_{I}=\frac{1}{2D},\\
		&\mathcal{V}^{\mathrm{pure}}_{C}=1+\frac{1}{D}.
	\end{align}
\end{subequations}
On the other hand, one can consider the case of a fully-mixed biphoton state, defined as
\begin{equation}
	\rho^{\text {mixed}}:=\frac{1}{D} \sum_{i=1}^{D} \ket{\Psi_i}\bra{\Psi_i},
\end{equation}
where $\ket{\Psi_i}\equiv\ket{1}_{a}\otimes\ket{1}_{b}$ is a pure separable biphoton state. Its purity is determined by a number of classical mixtures $D$ in the fully-mixed state. According to Eq.\ref{eq:Purity2009}, the purity and visibility of two-fold coincidences are  
\begin{subequations}
	\begin{align}	
		&\mathcal{P}^{\mathrm{mixed}}=\frac{2}{d}\\
		&\mathcal{V}_{C}^{\mathrm{mixed}}=\frac{4}{d}
	\end{align}
\end{subequations}

As $d\to\infty$, the corresponding statistical distribution of two-fold coincidences tends to a narrow Gaussian profile whereas that of pure maximally-entangled biphoton state converges to the exponential decay~\cite{Beenakker2009}.

\begin{figure}[htp]
	\centering
	\includegraphics[width=0.95\columnwidth]{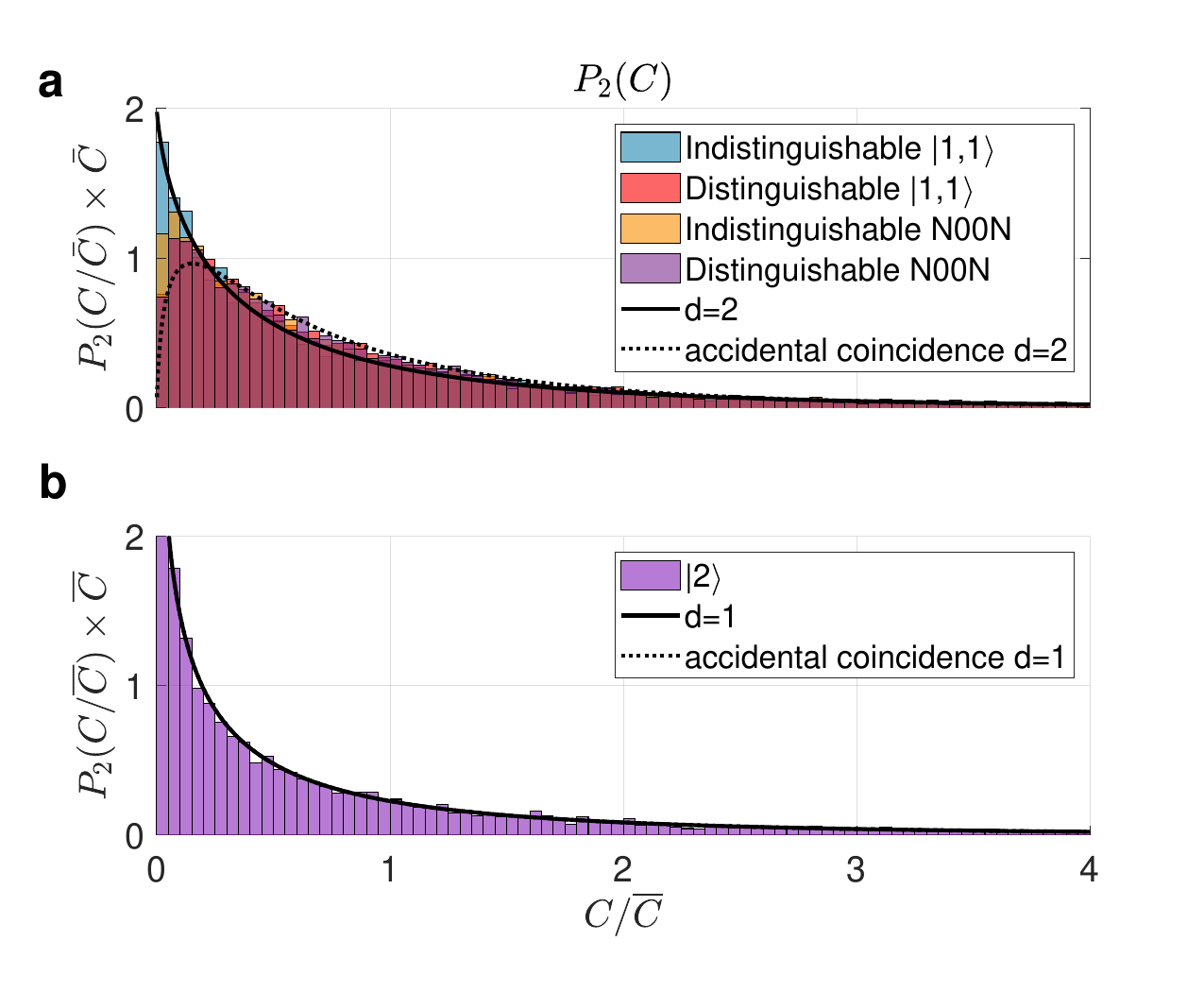}
	\caption{\textbf{Simulated statistical distributions of  two-fold coincidences}: (a) Two-mode states. The indistinguishable N00N state experiences the dephasing effect at the input of a random interferometer. (b) Two-photon Fock state. $P_{2}(R/\overline{R})$ has the same form as $P_{2}(C/\overline{C})$ when $d=1$.}
	\label{figPDFC_sim}
\end{figure}

In the case of separable biphoton states, we compare the theoretical prediction of $P_{2}\left(C/\overline{C}\right)$ in the presence of imperfect indistinguishability by means of the simulation. The result as shown in Fig.~\ref{figPDFC_sim}a indicates that the feature of quantum interference is essentially presented at $C/\overline{C}\rightarrow0$, which can be difficult to obverse in the experiment in the presence of noises. 

\subsection{Statistical distribution of accidental coincidences}
\label{SI_accidental_coincidences}
In order to take into account the contribution of accidental coincidences, we provide the data analysis of $R$. According to the fact that photocurrents $I_1$ and $I_2$ are two independent random variables with the PDF of Eq.\ref{eq:PDF_I}, due to no correlations presented in the random measurements (\ref{SI:Statistical analysis of random measurements}). We calculated the probability density function of accidental coincidences $R$, and it reads,
\begin{equation}
	\label{eq:PDF_ACC}
	P_{2}(\frac{R}{\overline{R}})=\frac{2}{\Gamma(d)^2}d^{2d} \left(\frac{R}{\overline{R}}\right)^{d-1} \mathcal{K}_{0}\left(2d \sqrt{\frac{R}{\overline{R}}}\right).
\end{equation}

\subsection{Statistical distribution of normalized second-order intensity correlation}
The statistical distributions of normalized second-order intensity correlation $g^{(2)}$ are calculated. In the case of indistinguishable monochromatic two-photon interference, $g^{(2)}_\text{Indis.}$ reads,
\begin{equation}
    \label{eq:g2Indis}
	g^{(2)}_\text{Indis.}=\eta_S\frac{|t_{ik}t_{jl}+t_{il}t_{jk}|^2}{ (|t_{ik}|^2 +|t_{il}|^2)(|t_{jk}|^2 +|t_{jl}|^2)},
\end{equation}
where $t_{ab}$ is the transmission amplitude from input $b$ to output $a$, and $\eta_S$ is the normalization related to the brightness of the light sources in the unit of coincidence per accidental count rates. In the case of the two-photon Fock state, this reduces to $g^{(2)}_\text{\ket{2}} =2\eta_S$ which is experimentally observed (Fig.~\ref{fig:Mainresults}a). 

\begin{figure}[htp]
	\centering
	\includegraphics[width=\columnwidth]{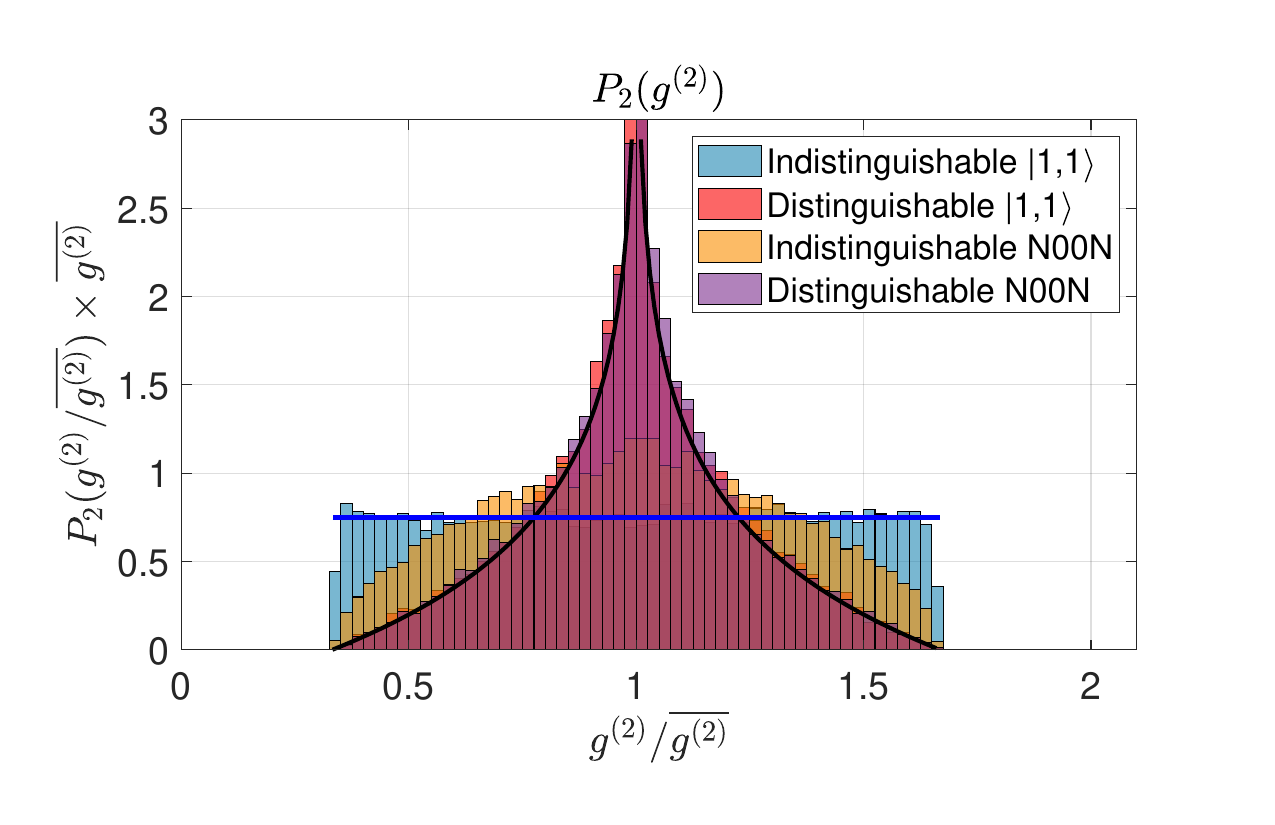}
	\caption{\textbf{Simulated statistical distributions of normalized second-order correlation}: Indistinguishable two-photon state follows the uniform distribution, whereas fully distinguishable states follow Eq.\ref{eq:PDFg2Dis}. The effect of dephasing in the case of the $N=2$ N00N state causes the reduction in $\mathcal{V}_{g^{(2)}}$. The simulation includes the presence of accidental coincidences.}
	\label{figPDFg2nor_sim}
\end{figure}

For indistinguishable monochromatic two-photon interference, we found numerically that the PDF is expected to be: 
\begin{equation}
    \label{eq:PDFg2Indis}
	P(g^{(2)}_\text{Indis.})=\frac{1}{2\overline{g^{(2)}}},
\end{equation}
where $g^{(2)}_\text{Indis.}\in[0,2\overline{g^{(2)}}]$. And for distinguishable monochromatic two-photon interference, $g^{(2)}_\text{Dis.}$ can be expressed as,
\begin{equation}
\label{eq:g2Dis}
	g^{(2)}_\text{Dis.}={\eta_S}\frac{|t_{ik}t_{jl}|^2+|t_{il}t_{jk}|^2}{(|t_{ik}|^2 +|t_{il}|^2)(|t_{jk}|^2 +|t_{jl}|^2)}.
\end{equation}
We found empirically that the PDF in Eq.\ref{eq:PDFg2Dis} is well described by the distribution,
\begin{equation}
    \label{eq:PDFg2Dis}
	P(g^{(2)}_\text{Dis.})= -\frac{\log{|g^{(2)}_\text{Dis.}/\overline{g^{(2)}}-1|}}{\pi\overline{g^{(2)}}},
\end{equation}
It is important to note that the PDF presents the singularity at $g^{(2)}_\text{Dis.}=\overline{g^{(2)}}$. The simulated results are shown in Fig.\ref{figPDFg2nor_sim}.
\onecolumngrid
\clearpage
\newpage
\newpage
\section{SPDC source}
\label{SI:SPDC_simulation}

\begin{figure*}[htp]
	\centering
	\includegraphics[width=0.85\linewidth]{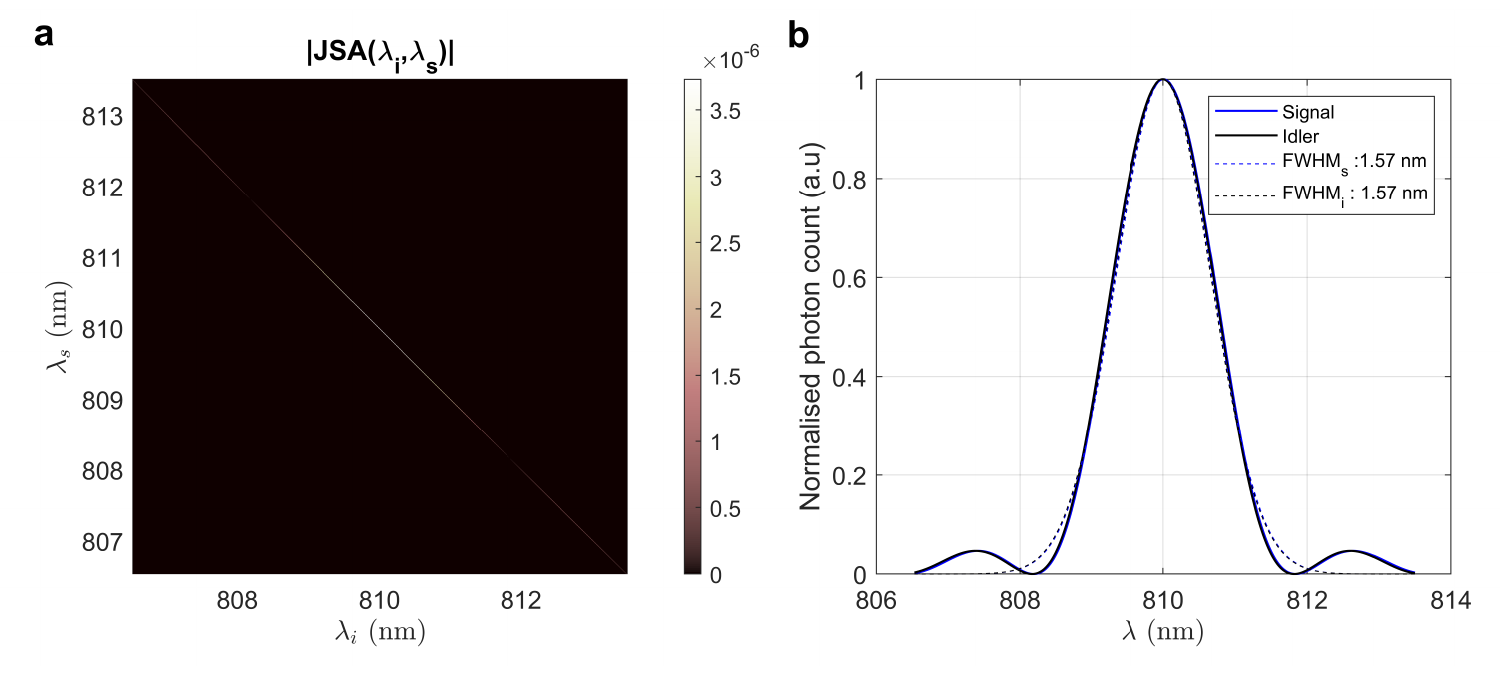}
	\caption{\textbf{Simulated JSA of the SPDC source}: (a) $|\text{JSA}(\lambda_i,\lambda_s)|$ (b) Spectral profiles for signal and idler photons}
	\label{figSPDC}
\end{figure*}

The degenerated spatially separable down-conversed biphoton state generated from ppKTP crystal is described as:
\begin{equation}
	\label{eq:twophotonstate}
	|\Psi\rangle=\int\int\Psi\left(\omega_{s}, \omega_{i}\right) \hat{a}_{s,H}^{\dagger}\left(\omega_{s}\right) \hat{a}_{i,V}^{\dagger}\left(\omega_{i}\right) d \omega_{s} d \omega_{i}|0\rangle,
\end{equation}
where $\Psi\left(\omega_{s}, \omega_{i}\right)$ is the joint spectral amplitude (JSA). The simulated JSA is provided in Fig.~\ref{figSPDC} to support the presence of anti-correlation in the spectral domain. 

\section{Statistical analysis of random measurements}
\label{SI:Statistical analysis of random measurements}

\begin{figure*}[htp]
	\centering
	\includegraphics[width=0.8\linewidth]{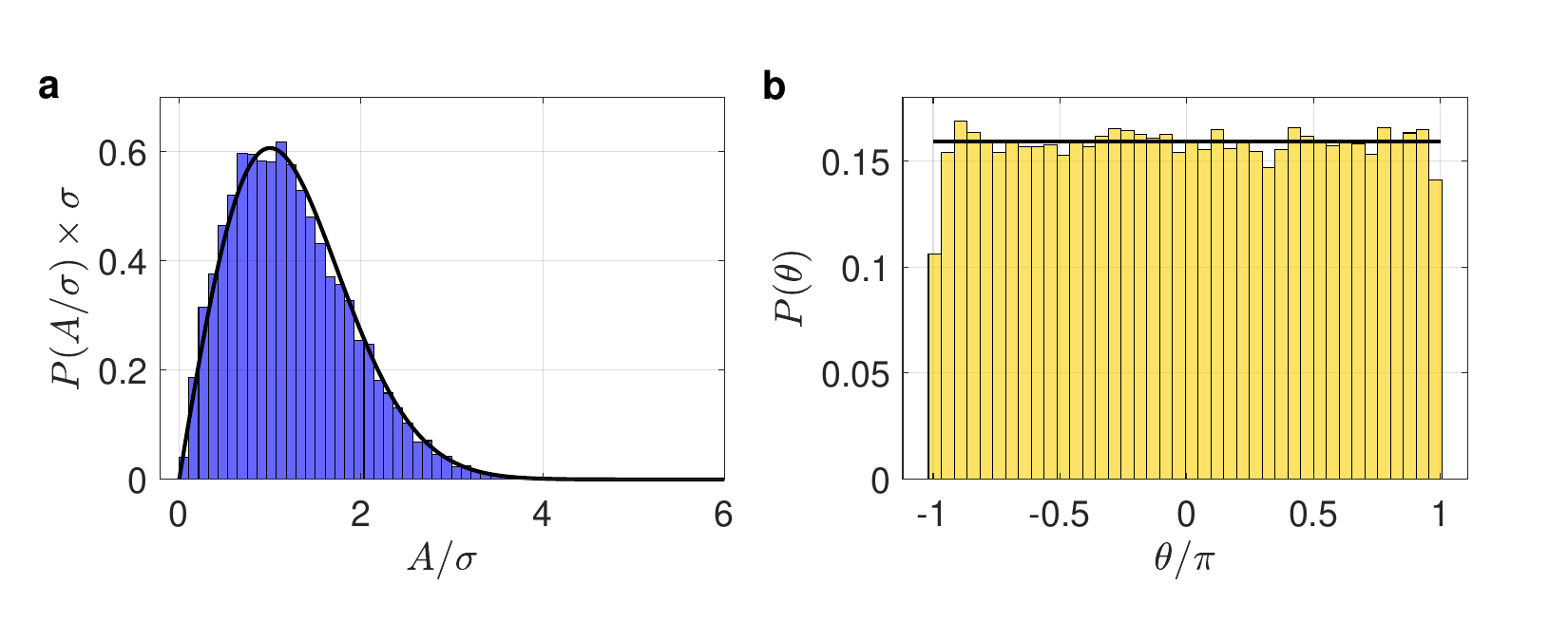}
	\caption{Statistical distribution of random transformations: (a) probability density functions of the amplitude $A$ and (b) of the phase $\theta$.}
	\label{figPDFtheta}
\end{figure*}

The statistical analysis of the random measurements implemented by the SLM and MMF on two truncated output modes is provided. As shown in Fig.~\ref{figPDFtheta}, the statistical distributions of amplitude and phase of random transformations in the case of 55-cm long MMF are shown to be consistent with i.i.d. complex Gaussian random variables.

\section{Experimental parameters and status of ground-truth light sources}
\label{SI:Experimental parameters and status of ground-truth light sources}
The experimental setting and status of the ground-truth states are presented in Table~\ref{table:Expdetail}. The overall measurement time was a few days per ground-truth state. To verify the stability of the sources and of the optical setup, we kept tracking the mean photocurrent at the two outputs and the mean and variance of the normalized second-order intensity correlation. Furthermore, the visibilities of HOM interference were measured before starting and after finishing the experiments in order to ensure stability in the cases of the SPDC source.

\begin{table}[htp]
	\centering
	\caption{Experimental setting and status of ground-truth light sources. $\mathcal{N}$ is the number of random measurements. $V_\text{before(after)}$ is a visibility of two-photon interference before and after performing an experiment. $T$ is the integration time of the detector for each measurement setting, $\delta$ is the temporal optical delay, and $l_c$ is the two-photon coherence length.}
	\label{table:Expdetail}
	\vspace{3mm}
	\begin{tabular}{@{}lccccc@{}}
		\toprule \midrule
		\multicolumn{1}{l}{State} & $T$ (s)  & $\mathcal{N}$ & $V_\text{before}$ & $V_\text{after}$ & $\Delta\lambda$ (nm) \\ \midrule \midrule
		\textbf{Monochromatic case} 	& &      &      &      &	  \\
		Incoherent source ($d=2$)  		& 5 & 10000 & -    &  -     &  2.35\\
		Biphoton state ($\delta=0$) 	& 15 & 10000 & 0.95 & 0.94 & 1.54  	\\ 
		Biphoton state ($\delta>l_c$)   & 15 & 10000 & 0.94 & 0.74 & 1.53	\\
		N=2 N00N state 				& 15 & 10000 & 0.98 & 0.98 & 1.65 		\\
		Two-photon state $\ket{2_V}$  					& 15 & 10000 & 0.98 & 0.82 & 1.53	\\
		Single-photon state $\ket{1_V}$ & 15 & 10000 & 0.98 & 0.98 & 1.54  	\\
		\textbf{Polychromatic case}  	& &      &      &          & \\
		Incoherent dispersive source    & 15 & 4004   &  -   & -    & 2.35\\
		Biphoton state ($\delta=0$)    	& 15 & 10000 & 0.94& 0.73  &  1.60  \\
		Biphoton state ($\delta>l_c$)  	& 15 & 10000  & 0.82 & 0.92 & 1.52   \\
		\bottomrule
	\end{tabular}
\end{table}

\section{State properties measured by random measurements}
\label{SI:Purity and dimensionality of biphoton states}
The properties of states are provided in Table~\ref{table:PandDtableApp}. The purity $\mathcal{P}$ of ground-truth monochromatic biphoton states is calculated from the visibilities of intensity $\mathcal{V}_I$, and of two-fold coincidences $\mathcal{V}_C$ using Eq.\ref{eq:Purity2009}. The number of occupied modes $d$ are obtained by $1/\mathcal{V}_I$.
\begin{table}[htp]
\begin{threeparttable}
	\centering
	\caption{Visibilities and estimated properties of ground-truth states.}
	\label{table:PandDtableApp}
	\vspace{3mm}
	\begin{tabular}{@{}lcccccc@{}}
		\toprule\midrule
		\multicolumn{1}{l}{State} & $\mathcal{V}_I$ & $\mathcal{V}_C$ & $\mathcal{V}_C$\tnote{*} & $\mathcal{P}$ & $d$ & $\mathcal{V}_{g^{(2)}}$ \\ 	\midrule \midrule
		\textbf{Monochromatic case}  &    &    &    &    &    & \\
		Incoherent source ($d=2$)  & $0.44\pm0.02$ &  $1.14\pm0.06$\tnote{b} & -  &  -   &  $2.27\pm0.07$ & $0.021\pm0.003$ \\
		Biphoton state ($\delta=0$) & $0.46\pm0.02$  & $1.46\pm0.09$  & $1.38\pm0.09$         & $0.45\pm0.11$     & $2.16\pm0.06$ & $0.178\pm0.009$		\\ 
		Biphoton state ($\delta>l_c$)  & $0.50\pm0.02$  & $1.49\pm0.09$   & $1.34\pm0.09$     & $0.35\pm0.11$   &$2.02\pm0.05$	& $0.096\pm0.007$	\\
		N=2 N00N state  & $0.45\pm0.02$  & $1.43\pm0.09$  & $1.27\pm0.08$  & $0.38\pm0.08$   &$2.24\pm0.01$ &	$0.127\pm0.007$	\\
		Two-photon state $\ket{2_V}$  & $0.80\pm0.04$\tnote{a}  & $2.52\pm0.17$  & $2.39\pm0.15$     & $0.69\pm0.22$   &$1.23\pm0.09$ &  $0.105\pm0.011$	\\
		Single-photon state $\ket{1_V}$  & $0.81\pm0.04$\tnote{a}  & $2.37\pm0.02$\tnote{b} &  -    & -   &$1.23\pm0.07$  &  $0.402\pm0.048$	\\
		\textbf{Polychromatic case}  &    &    &    &    &    & \\
		Incoherent dispersive source  &  $0.048\pm0.008$    &  $0.09\pm0.01$\tnote{b} & -   & -   & $20\pm1$ & $0.0013\pm0.0012$  \\
		Biphoton state ($\delta=0$)    &  $0.068\pm0.006$   & $0.27\pm0.02$  &  $0.26\pm0.02$   &  $0.12\pm0.02$   & $14.7\pm0.2$ & $0.074\pm0.006$   \\
		Biphoton state ($\delta>l_c$)  & $0.073\pm0.006$    & $0.26\pm0.01$  &  $0.22\pm0.01$   &  $0.08\pm0.02$   & $13.8\pm0.3$ &  $0.049\pm0.005$  \\
		\bottomrule
	\end{tabular}
	\begin{tablenotes}\footnotesize
        \item[*] corrected for accidental coincidences.
        \item[a] By subtracting estimated dark counts, $\mathcal{V}_I$ values are corrected to $0.85\pm0.06$ and $0.9\pm0.08$ for $\ket{2_V}$ and $\ket{1_V}$, respectively and resulting accordingly in the corrected $d$ of $1.18\pm0.09$ and $1.11\pm0.10$. 
        \item[b] contribution of accidental coincidences and $\overline{g^{(2)}}=1$.
    \end{tablenotes}
    \end{threeparttable}
\end{table}

\section{Underestimation of the purity}
\label{sec:Estimation of the purity}
\begin{figure*}[htp]
	\centering
	\includegraphics[width=0.7\linewidth]{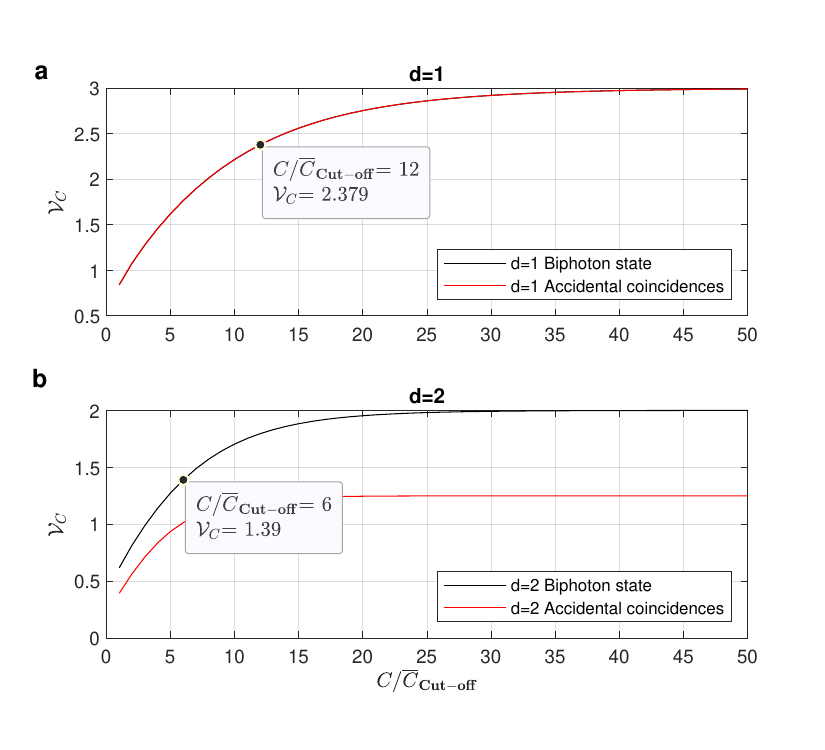}
	\caption{\textbf{Underestimation of $\mathcal{V}_C$ at different $C/\overline{C}_\text{Cut-off}$}: due to the lack of rare events (a) $d=1$ (b) $d=2$}
	\label{fig:Vccutoff}
\end{figure*}

As reported in Table~\ref{table:PandDtableApp}, the estimation of purity is lower than the expected values of one. This mainly is because of the underestimation of $\mathcal{V}_C$. We attribute the error to two contributions: the influence of accidental coincidences and the lack of rare events originating from the long tail of the statistical distributions.
\begin{enumerate}
	\item The first is the influence of accidental coincidences since it modifies the distribution of two-fold coincidences to exhibit less probability of detecting coincidence counts close to zero. The experimental values of $\mathcal{V}_C$ reported for two-photon states in Table~\ref{table:PandDtableApp} after correction for expected accidental coincidences (${\mathcal{V}_C}^*$) is greater than the estimation of $\mathcal{V}_C$ of 1.25 from the PDF of accidental coincidences (Eq.\ref{eq:PDF_ACC}) at $d=2$. The calibrated values are, however, still underestimated which we attribute to the lack of rare events.
	
	\item The second contribution results from a long tail of the statistical distributions. As the distributions of two-fold coincidences and accidental coincidences return to be identical in the cases of $d$=1, we can thus neglect the influence of accidental coincidences on $\mathcal{V}_C$ and use the $d=1$ distributions to investigate the effect of the lack of rare events (Fig.~\ref{fig:Vccutoff}a). The $d=1$ distributions estimate $\mathcal{V}_{C}=3$, while experimentally we obtained 2.39 for both the two-photon Fock state $\ket{2}$ and the coherent state. These visibilities of two-fold coincidences are close to the value estimated from the PDFs with the maximum cut-off at $C/\overline{C}_\text{Cut-off}=12$ which implies the lack of rare events at higher $C/\overline{C}$. For $d$=2, the same contribution is found in Fig.~\ref{fig:Vccutoff}b. The estimated $\mathcal{V}_C$ using the PDF with the maximum cut-off at $C/\overline{C}_\text{Cut-off}=6$ is of $1.39$ which is close to the experimental values for monochromatic biphoton states (Table~\ref{table:PandDtableApp}). To obtain an accurate value of $\mathcal{V}_C$, hence the purity, the significant number of race events up to $C/\overline{C}_\text{Cut-off}>25$ are needed.
	
\end{enumerate}

\newpage
\section{Contribution of noises on the statistical distributions}
\label{SI:Noise}
\begin{figure*}[htp]
	\centering
	\includegraphics[width=0.7\linewidth]{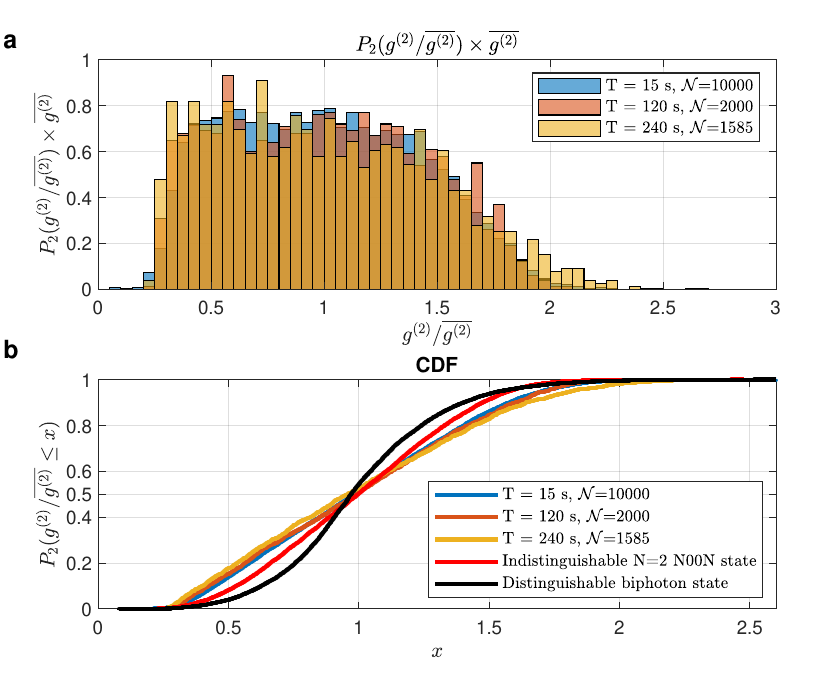}
	\caption{\textbf{Statistical distributions of normalized second-order correlation}: (a) Histograms of $g^{(2)}$ for the indistinguishable monochromatic biphoton state at different integration times $T$ and the number of measurements $\mathcal{N}$. The distributions present a broad and flattened feature compared to the distinguishable biphotons. (b) Cumulative distribution functions of $g^{(2)}$: comparing indistinguishable monochromatic biphoton state at three different integration times $T$ and the number of measurements $\mathcal{N}$ as labelled with that of the N00N state and of the distinguishable monochromatic biphoton state.}
	\label{figCh4compareTac}
\end{figure*}

We tested the influence of noises by varying the integration time $T$ and the number of measurements $\mathcal{N}$ on the statistical distributions in the case of indistinguishable monochromatic biphoton state. As shown in Fig.~\ref{figCh4compareTac}, we found that all statistical distributions in the case of integration times of 15, 120, and 240 s show the same broadened profiles with $\mathcal{V}_{g^{(2)}}$ of $0.178\pm 0.009$, $0.18\pm0.02$, and $0.22\pm0.03$, respectively. The results can be used to discriminate the indistinguishable biphoton state from the distinguishable state.

\newpage
\section{Statistical distributions of intensity}
\label{SI_Statistical distributions of intensity}
\begin{figure*}[htp]
	\centering
	\includegraphics[width=0.9\linewidth]{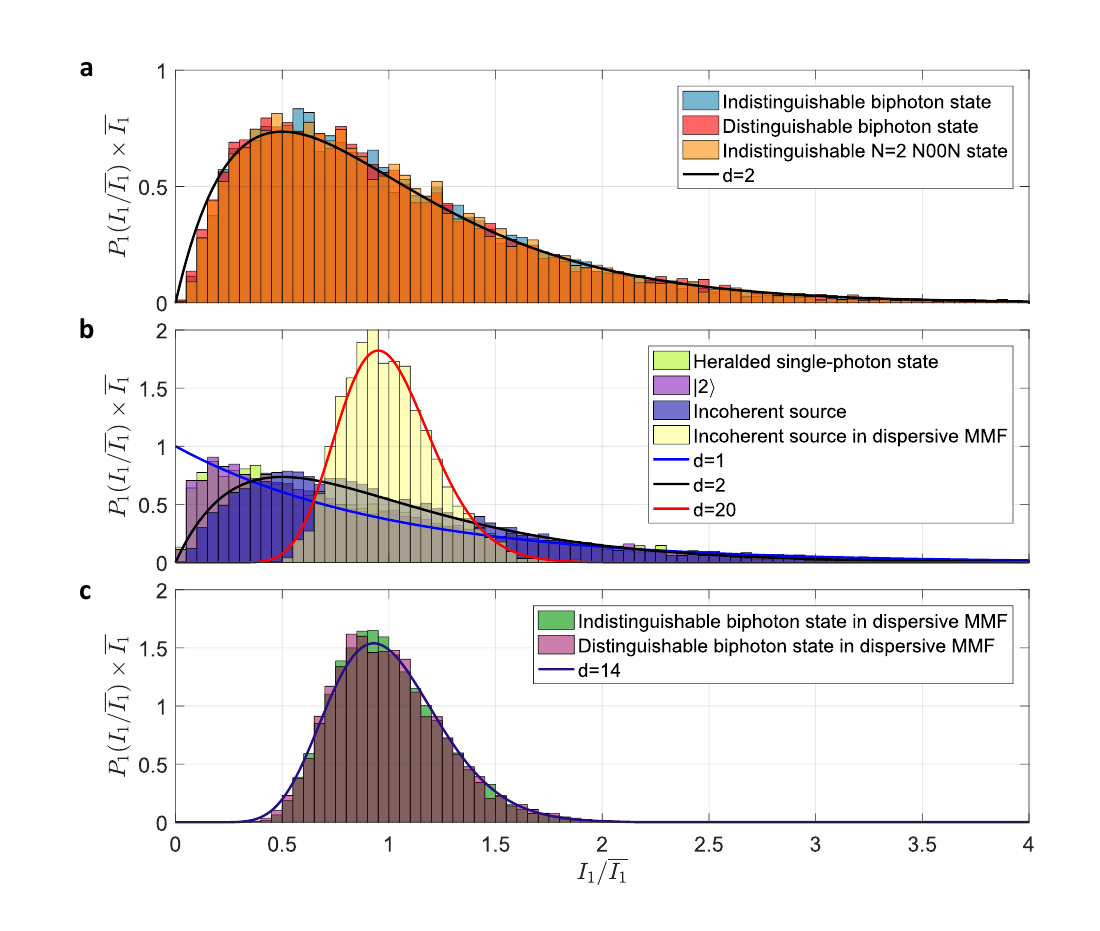}
	\caption{\textbf{Statistical distributions of intensity $P_1(I_1/\overline{I_1})$}: (a) Indistinguishable biphoton state (blue), distinguishable biphoton state (red), N=2 N00N state (orange) have the same distribution with $d=2$. (b) Heralded single-photon state (light green) and two-photon state (magenta) have the same exponential decay $d=1$. The distributions exhibit less probability of detecting intensity close to zeros due to the presence of dark counts. The incoherent source (dark blue) in a 55-cm MMF and incoherent source (light yellow) in a 25-m MMF have different distributions with $d=2$ and $d=20$. (c) Indistinguishable (green) and distinguishable (light magenta) biphoton states propagating through the dispersive 25-m MMF have the same distribution $d\approx14$ $(d=14.7, d=13.8)$.}
	\label{fig:PDF_I1}
\end{figure*}

The statistical distributions of intensity speckles, as presented in Fig.~\ref{fig:PDF_I1}, provide the knowledge of the number of occupied modes for all ground-truth states. The same distributions are observed for the intensity on the second detection $I_2$ (the data are not shown).

\end{document}